\documentclass[twocolumn,showpacs,preprintnumbers,amsmath,amssymb]{revtex4}

\usepackage{subfigure}
\usepackage[dvips]{graphicx}
\usepackage{dcolumn}
\usepackage{bm}
 \newcommand{\eq}{\begin{equation}}
 \newcommand{\eqq}{\end{equation}}

\begin{document}


\title{Dynamic model and stationary shapes of fluid vesicles}

\author{F. \surname{Campelo}}
 \email{campelo@ecm.ub.es}
\author{A. \surname{Hern\'{a}ndez--Machado}}%
\affiliation{Departament d'Estructura i Constituents de la Mat\`{e}ria,\\
Facultat de F\'{\i}sica, Universitat de Barcelona\\
  Diagonal 647, E-08028 Barcelona, Spain}

\date{\today}

\begin{abstract}

A phase-field model that takes into account the bending energy of fluid vesicles is presented. The Canham-Helfrich model is derived in the sharp-interface limit. A dynamic equation for the phase-field has been solved numerically to find stationary shapes of vesicles with different topologies and the dynamic evolution towards them. The results are in agreement with those found by minimization of the Canham-Helfrich free energy. This fact shows that our phase-field model could be applied to more complex problems of instabilities.

\end{abstract}

\maketitle

\section{\label{sec:introduction}Introduction}

Biological membranes have been of wide interest to biologists, chemists and physicists for many years (cfr. \cite{edidin03} for a historical review). They are the frontier which defines cells and most of their internal organelles, and they act as a boundary between internal cellular organization and its surrounding medium. Membranes are composed of several kinds of lipids (phospholipids, cholesterol and glycolipids), which are self-assembled in a fluid bilayer, and by membrane proteins which are anchored on it \cite{alberts,kozlov06}. From the molecular point of view, biomembranes are extremely complex. However, there seems to be a universal construction principle common to all actual membranes, which is the presence of a fluid lipid bilayer through which proteins can diffuse. Vesicles are model closed membranes consisting of one or several different kinds of lipids \cite{seifert97,safran99}. They have therefore been studied to get an idea of the main physical properties of actual biomembranes \cite{lipowskysackmann}, such as red blood cells's \cite{alberts}.

Recently several experimental results have been reported on dynamic instabilities in membranes, such as pearling~\cite{tsafrir01,tsafrirphd}, budding and tubulation~\cite{tsafrir03,tsafrirphd}. In these experiments shape instabilities are induced by the insertion of a certain concentration (locally or globally) of an amphiphilic polymer (which mimicks the proteins within the biomembranes) in the outer leaflet of the bilayer \cite{ringsdorf88}. 

Dynamic models of vesicles in shear flow \cite{kraus96,misbah03,noguchi04,noguchi05}, domain growth in multicomponent membranes and bud formation \cite{taniguchi96,sunilkumar01,sunilkumar04,sens04}, and pearling \cite{goldstein96} have been studied theoretically. The effect of proteins included in lipidic membranes is also the subject of theoretical studies \cite{prost96,benamarphysa04,kozlov06}. 

Phase-field models (or diffuse interface models) can be thought as mathematical tools to study complex interfacial problems, such as free boundary problems \cite{langer86}. Phase-field models are mesoscopic models of the Ginzburg-Landau type, which disregard microscopic details. Such models have been widely used before in different interfacial problems such as solidification and the Saffman-Taylor problem \cite{gonzalezcinca04} and roughening \cite{alava04}. Most of these phase-field models describe the effect of surface tension, but do not deal with bending energies.

Biben and Misbah studied the tumbling transition of a vesicle under shear flow \cite{misbah03} using a phase-field-like model. Their model couples the bending energy with the velocity field as a force in the two dimensional hydrodynamic equation (in the Stokes limit). Du \textit{et al.} used a phase-field model \cite{duliuwang04,duliuwang06} based on the free energies for surfactant films found in the works of Gompper and Schick \cite{gompperschick90,gompperschick}. They applied their model to minimize the free energy to find stationary shapes, using a standard global Lagrange formalism  to deal with both global area and volume constraints. The minimization was performed by a gradient flow equation, through a non-conserved purely relaxational dynamics.


In this paper we have derived a phase-field model for the bending energy of fluid vesicles. The membrane is considered as a mathematical interface between two \textit{phases}, the inner fluid and the outer fluid. In this kind of models there is no need to track the interface during the dynamic evolution, which is one of the main problems in membrane dynamics \cite{kraus96}. Our equations are continuous in the whole domain, and the interface is located by the level-set of the phase-field, i.e. the region of rapid variation of the phase-field. The free energy functional associated with this model reduces to the Canham-Helfrich bending energy of the lipid bilayer \cite{{canham70},{helfrich73}} in the so-called sharp interface limit, when interface width goes to zero. In addition, phase-field models are dynamic models, so we are capable with our model to study dynamic properties of vesicles, such as relaxation towards stationary shapes. The fact that we find the correct stationary shapes shows that our free energy functional deals correctly with bending energies.

This paper is organized as follows: In Sec.~\ref{sec:model} we propose a phase-field model to take into account the bending energy of fluid vesicles, and we prove that it is equivalent to the Canham-Helfrich model. An effective free energy functional is written down in the presence of geometric constraints, and the dynamic equation is worked out. Details of the numerical integration are presented in Sec.~\ref{sec:num.int}. The stationary shapes for spherical and non-spherical topologies are presented in Sec.~\ref{sec:results} together with some features of dynamic evolution. A shape diagram is presented in Sec.~\ref{sec:discussion} for vesicles with spherical topology. Discussion of the possible applications and extensions of this model, and main conclusions are found at the end of the paper in Sec.~\ref{sec:conclusions}.

\section{\label{sec:model}Model}

\subsection{\label{sec:canham-helfrich}Canham--Helfrich free energy}

A mathematical model describing the geometric properties of lipidic membranes was introduced by Canham~\cite{canham70} and Helfrich~\cite{helfrich73} to describe interfaces governed by bending energy. The membrane is considered to be a two-dimensional surface embedded in Euclidean three-di\-men\-sional space. Such a surface has two curvature invariants: the mean and the Gaussian curvatures. The so-called Can\-ham-Helfrich local bending energy is thus an expansion in terms of these invariants. For membranes with a possible assymmetry between the two layers, the spontaneous curvature model \cite{seifert97} reads
\begin{equation}
f_{\,\mathrm{C-H},\ \mathrm{sc}}=\frac{\kappa}{2}\left(2 H-c_0\right)^2+\kappa_G K,
\label{localbending}
\end{equation}
where $\kappa$ and $\kappa_\mathrm{G}$ are two elastic constants: the bending rigidity, and the Gaussian bending rigidity, respectively; and $c_0$ is the spontaneous curvature.
Due to the Gauss--Bonnet theorem, the Gaussian curvature term (the last term in Eq.~(\ref{localbending}))
integrated over a closed surface is a topological invariant. Since we are not concerned with studying topological changes here, this term will be a constant factor in the total free energy, so it does not need to be considered. Therefore the bending energy reduces (after integration over the whole vesicle membrane surface) to
\eq\label{globalbending_sc}
F_{\,\mathrm{C-H},\ \mathrm{sc}}=\frac{\kappa}{2}\int_{\Gamma}\left(2H-c_0\right)^2\mathrm{d}\bm{s},
\eqq
where $\Gamma$ is the membrane surface.
The special case where $c_0=0$ in Eq.~(\ref{globalbending_sc}) is the minimal model
\eq\label{globalbending}
F_{\,\mathrm{C-H}}=\frac{\kappa}{2}\int_{\Gamma}\left(2H\right)^2\mathrm{d}\bm{s},
\eqq
where no asymmetry between the two layers of the lipidic membrane is considered.

\subsection{\label{sec:themodel}Phase--field implementation}

\subsubsection{Minimal Model}

In order to obtain our phase-field model, we will start with an \textit{ansatz} for the free energy functional, which will be shown to be equivalent, under certain limits, to the Canham-Helfrich Hamiltonian Eq.~(\ref{globalbending}), with no spontaneous curvature. This free energy functional is
\eq
\mathcal{F}[\phi]=\int_{\Omega}{\Phi^2[\phi\,]\ \mathrm{d}\bm{x}},
\label{ansatz}
\eqq
where the function $\phi$ is the so-called phase-field, and
\eq\label{chem.pot}
\Phi[\phi(\bm{x})]=-\phi+\phi^3-\epsilon^2\bm{\nabla}^2\phi(\bm{x}),
\eqq
where $\epsilon$ is a small parameter related to the interface width. Note that the free energy density functional in Eq.~(\ref{ansatz}), $\Phi^2$, is nothing else other than the square of the chemical potential (the functional derivative of the free energy) associated with the Cahn-Hilliard problem \cite{cahnhilliard58}.

The minimum of the free energy Eq.~(\ref{ansatz}), with no constraints, is obtained by setting Eq.~(\ref{chem.pot}) equal to 0. In one dimension, this leads to the tanh-like solution $\phi(x)=\tanh{\left(\frac{x}{\sqrt{2}\epsilon}\right)}$, given the boundary conditions $\phi(\pm \infty)=\pm 1$. The boundary conditions in three dimensions are that the phase-field at infinity is $\phi=-1$, which is the value for the stable phase of the outside bulk. 

Phase fields are regular functions, so they can be written in terms of the signed distance to the interface, the set $\phi=0$, in units of $\epsilon$,
\eq\label{phi.f}
\phi(\bm{x})=f\big(D(\bm{x})\big),
\eqq
where $D(\bm{x})=d(\bm{x})/\epsilon$, and $d(\bm{x})$ is the signed distance of the point $\bm{x}$ to the surface $\Gamma$.
We should plug this into Eq.~(\ref{ansatz}). However, before doing this, we will explicitly work out the Laplacian term
\begin{eqnarray}\label{nabla2}
\bm{\nabla}^2\phi &=& \bm{\nabla}\Big(\bm{\nabla} f\big(D(\bm{x})\big)\Big)=\bm{\nabla}\left(f'\big(D(\bm{x})\big)\frac{\bm{\nabla} d(\bm{x})}{\epsilon}\right) \nonumber \\
&=&f''\big(D(\bm{x})\big)\left(\frac{\bm{\nabla} d(\bm{x})}{\epsilon}\right)^{\!\! 2} + f'\big(D(\bm{x})\big)\frac{\bm{\nabla}^2 d(\bm{x})}{\epsilon}.
\end{eqnarray}
We now consider that our surface is regular enough to let the distance be just the position vector (plus some constant vector, which will be irrelevant when differentiated). Thus,
\eq\label{grad.dist}
\bm{\nabla} d(\bm{x})=\hat{n},
\eqq
is a unit vector normal to the interface. If we insert Eq.~(\ref{grad.dist}) into Eq.~(\ref{nabla2}), we obtain
\eq\label{nabla2.}
\bm{\nabla}^2\phi =\frac{1}{\epsilon ^2} f''\big(D(\bm{x})\big) + \frac{1}{\epsilon} f'\big(D(\bm{x})\big)\bm{\nabla}^2 d(\bm{x}).
\eqq
Thus, taking the result from Eq.~(\ref{nabla2.}), we see that Eq.~(\ref{ansatz}) reduces to
\eq\label{wf}
\mathcal{F}(f)= \int_{\Omega} \left[\big(f''-(f^2-1)f\big) + \epsilon f'\big(D(\bm{x})\big)\bm{\nabla} ^2 d(\bm{x}) \right] ^2 \mathrm{d}\bm{x}.
\eqq
This is the free energy functional that we would like to minimize dynamically under certain constraints.

In the sharp-interface limit ($\epsilon \to 0$) and, in order to minimize the free energy functional, Eq.~(\ref{wf}), we obtain
\eq\label{diff}
f''\big(D(\bm{x})\big)-\left(f^2\big(D(\bm{x})\big)-1\right)f\big(D(\bm{x})\big)= 0.
\eqq
Therefore we obtain the interfacial solution
\eq\label{tanh}
f\left(D(\bm{x})\right)=\tanh{\!\left(\frac{d(\bm{x})}{\epsilon\sqrt{2}}\right)},
\eqq
where we can see that $\epsilon\sqrt{2}$ is the width of the interface. Thus, Eq.~(\ref{wf}) reduces to
\eq\label{wff}
\mathcal{F}(f)= \int_{\Omega}\left|\epsilon f'(D(\bm{x}))\bm{\nabla} ^2 d(\bm{x})\right| ^2 \mathrm{d}\bm{x}.
\eqq
From Eq.~(\ref{tanh}), we can compute
\eq f'\left(D(\bm{x})\right)=\frac{1}{\sqrt{2}}\mathrm{sech}^2\left(\frac{d(\bm{x})}{\epsilon\sqrt{2}}\right).
\eqq
Using the fact that
\eq\label{delta.repr}
\lim_{\epsilon \to 0}\left\{\frac{3}{4\sqrt{2}\epsilon}\ \mathrm{sech}^4\left(\frac{d(\bm{x})}{\epsilon \sqrt{2}}\right)\right\}=\delta\left(d(\bm{x})\right),
\eqq
we can write the integral over the volume $\Omega$ in Eq.~(\ref{wff}), as an integral over the surface $\Gamma$
\eq
\mathcal{F}(d)= \frac{2\sqrt{2}}{3}\ \epsilon^3 \int_{\Gamma}\left|\bm{\nabla}^2 d(\bm{x})\right| ^2 \mathrm{d}\bm{s}.
\eqq

Using the fact that (cfr. e.g. \cite{docarmo76,safran94})
\eq\label{hnabla2}
H=\frac{1}{2}\bm{\nabla} ^2 d(\bm{x}),
\eqq
we can see that the free energy (\ref{ansatz}) is, up to a known proportionality constant, the same as the Canham-Helfrich minimal model Eq.~(\ref{globalbending}). We may thus employ the phase-field dependent functional to implement the bending energy in a vesicle membrane. Since the proportionality constant is positive, it just redefines the energy scale. It is thus exactly equivalent to minimizing our free energy functional instead of the curvature dependent one.

\subsubsection{Spontaneous Curvature}

It is possible to extend the results of the last section in order to include a spontaneous curvature, as in Eq.~(\ref{globalbending_sc}). The new \emph{ansatz} for the free energy functional will be the same as in Eq.~(\ref{ansatz}), but with an additional term in the free energy density
\eq
\mathcal{F}_{\mathrm{sc}}[\phi]=\int_{\Omega}{\Phi^2_{\mathrm{sc}}[\phi\,]\ \mathrm{d}\bm{x}},
\label{ansatz_sc}
\eqq
where
\eq
\Phi_{\mathrm{sc}}[\phi(\bm{x})]=\Phi[\phi(\bm{x})]+\epsilon\,C_0(\bm{x})\,(1-\phi^2),
\eqq
where $C_0(\bm{x})$ is, as we show below, related to the spontaneous curvature which, in principle, can be position-dependent (or even $\phi$-dependent). The proof that Eq.~\ref{ansatz_sc} is equivalent to the SC model Eq.~(\ref{globalbending_sc}) is analogous to that shown in Sec.~\ref{sec:themodel}. Therefore, Eq.~(\ref{wf}) will read now as
\begin{eqnarray}
\mathcal{F}_{\mathrm{sc}}(f)&=& \int_{\Omega} \left[\big(f''-(f^2-1)f\big) + \epsilon\left( f'\,\bm{\nabla} ^2 d(\bm{x})\right.\right.\nonumber\\
&&\left.\left.-C_0 (1-f^2)\right) \right] ^2 \mathrm{d}\bm{x}.
\end{eqnarray}
The order $\epsilon$ term can be rewritten as
\eq
f'\,\bm{\nabla} ^2 d(\bm{x})-C_0 (1-f^2)=\left(\bm{\nabla} ^2 d(\bm{x})-\sqrt{2}C_0\right)\,f',
\eqq
where we used the fact that $1-f^2=\sqrt{2} f'$, which comes directly from Eq.~(\ref{tanh}). We finally reach
\eq
\mathcal{F}_{\mathrm{sc}}(d)= \frac{2\sqrt{2}}{3}\ \epsilon^3 \int_{\Gamma}\left|\bm{\nabla}^2 d(\bm{x})-\sqrt{2}C_0\right| ^2 \mathrm{d}\bm{s},
\eqq
which, identifying $\sqrt{2}C_0 = c_0$, is equivalent (again up to a constant) to the spontaneous curvature model Eq.~(\ref{globalbending_sc}).

\subsection{\label{sec:geom.constraints}Geometrical constraints}

Vesicle shapes are subject to certain geometrical constraints. Their enclosed volume and their surface area should remain constant. We will need to implement these constraints in our phase-field model.

\subsubsection{\label{sec:surface}Local surface area}

Lipidic membranes are in a fluid phase at physiological temperatures. The solubility of membrane lipids is extremely low, which implies no relevant exchange of material between the membrane and the surrounding media. In addition, the membrane can be considered to be locally incompressible. These two facts provide us with a constraint for the local area of the vesicle, which remains fixed.

We have implemented this constraint in our phase-field model via a Lagrange multiplier function coupled with the surface area in the free energy functional. Thus, we define an effective free energy functional
\eq
\mathcal{F}_{\mathrm{eff}}[\phi]=\mathcal{F}[\phi]+\int_{\Omega}{\sigma(\bm{x}) a[\phi]\mathrm{d}\bm{x}},
\label{effective0}
\eqq
where $\mathcal{F}[\phi]$ is given by Eq.~(\ref{ansatz}), $\sigma$ is a Lagrange multiplier (interpreted as a surface tension), and $a[\phi]$ is the local surface area functional,
\eq\label{area}
a(\bm{x})=\delta\left(d(\bm{x})\right),
\eqq
which we rewrite in terms of the parameter $\epsilon$ using the representation of the delta, Eq.~(\ref{delta.repr}),
\eq\label{area.eps}
a[\phi]=\frac{3}{2\sqrt{2}}\ \epsilon\left|\bm{\nabla} \phi\right|^2,
\eqq
where we have used the fact that $\epsilon\bm{\nabla}\phi(\bm{x})=f'(D(\bm{x}))\,\hat{n}$, and, from Eq.~(\ref{delta.repr}), that $(f'(D(\mathbf{x})))^2\sim \delta(d(\mathbf{x}))$.
Using Eq.~(\ref{area}), i.e. Eq.~(\ref{area.eps}) in the sharp-interface limit, the last expression integrated over the whole domain $\Omega$ is equivalent to the surface area of the vesicle,
\eq\label{globalarea}
\int_{\Omega}{a[\phi]\mathrm{d}\bm{x}}=\int_{\Gamma}\mathrm{d}\bm{s}.
\eqq

\subsubsection{\label{sec:volume}Enclosed volume}

The other geometrical property of the vesicle that requires a constraint is its enclosed volume. Biological membranes are permeable to water, but not to, e.g., large ions (on the time scales we are interested in) \cite{alberts}. This means that any transfer of water through the membrane would create an osmotic pressure which cannot be counterbalanced by the relatively much weaker bending energy,\cite{helfrich73}. Therefore, the concentration of osmotically active molecules fixes the inner volume of the vesicle. The usual way to implement this condition in the free energy is, as done before with the surface area constraint, to introduce a Lagrange multiplier coupled with the volume term, which ensures its conservation.

Thus, dynamics are introduced such that the vesicle volume is conserved through time. Model-B-like dynamics is therefore used
\eq\label{cons.dyn}
\frac{\partial \phi}{\partial t}=\bm{\nabla}^2\left(\frac{\delta\mathcal{F}_{\mathrm{eff}}}{\delta\phi}\right).
\eqq
This dynamic equation ensures that $\int_{\Omega}{\phi(\bm{x})\mathrm{d}\bm{x}}$ is constant in time. This integral is equal to the difference of the inner and outer volumes ($\phi$ takes its stable values $+1$ and $-1$ inside and outside the vesicle, respectively), as $\epsilon \to 0$. As the sum of the inner and outer volumes is the volume of the space $\Omega$ (which is constant), then we can write the inner volume as
\eq
V_{\mathrm{inn}}=\frac{1}{2}\left(V(\Omega)+\int_{\Omega}{\phi(\bm{x})\mathrm{d}\bm{x}}\right),
\eqq
where it can be seen that model-B dynamics ensures conservation of the inner volume through the dynamic evolution.

\subsection{\label{sec:dyn.eqn}Dynamic equation}

The dynamic relaxation towards free energy minima is achieved in our model by conserved relaxation dynamics, Eq.~(\ref{cons.dyn}). We could have introduced some homogeneous mobility, but this would have just changed the time scale. Performing the functional derivative in Eq.~(\ref{cons.dyn}) gives the dynamic equation
\begin{eqnarray}\label{dyn.eqn}
\frac{\partial \phi}{\partial t}&=&2\bm{\nabla}^2\Big\{(3\phi^2-1)\Phi[\phi]-\epsilon^2\bm{\nabla}^2\Phi[\phi]+\epsilon^2\bar{\sigma}(\bm{x})\bm{\nabla}^2 \phi\nonumber\\
&+&\epsilon^2 \bm{\nabla}\bar{\sigma}(\bm{x})\cdot\bm{\nabla}\phi\Big\},
\end{eqnarray}
where we have defined $\bar{\sigma}$ as
\eq\label{lagr.mult}
\bar{\sigma}(\bm{x})=\frac{\sqrt{2}}{3 \epsilon}\sigma(\bm{x}).
\eqq

Using this kind of dynamics, local conservation of the inner volume of the vesicle is achieved in a natural way, unlike Ref.~\cite{duliuwang06} which uses a purely relaxational dynamics with no direct conservation of the inner volume of the vesicle.

\section{\label{sec:num.int}Numerical Integration}

The partial differential equation (\ref{dyn.eqn}) is highly non-linear (notice, e.g., the coupling between the field $\phi^2$ and the functional $\Phi[\phi]$). We thus need to use a numerical method to solve this equation in order to study dynamic relaxation towards stationary vesicle shapes.

The discretization algorithms used are second-order finite differences for the spatial dependence, and an Euler scheme for the time dependence \cite{strikwerda}.

Our effective free energy functional (\ref{effective0}) explicitly contains a Lagrange multiplier. Therefore, we need to know the time evolution of the Lagrange multiplier. To do this, we have used a first order Lagrangian method to study how the multiplier evolves to its stationary value \cite{bertsekas}
\eq
\sigma^{k+1}(\bm{x})=\sigma^k(\bm{x})+\alpha \left(a[\phi^k(\bm{x})]-a_0(\bm{x})\right),
\eqq
where $\alpha>0$ is the stepsize, $k$ is the discretized time, and $a_0(\bm{x})$ is the fixed local surface area. Since we are not interested in the actual dynamics of the multiplier, our choice is justified because it does not change the dynamics of the phase-field, but it just keeps the surface area of the vesicle constant during the time evolution without altering the dynamics.

We have performed simulations on lattices of different sizes and equivalent shapes and evolutions were obtained. As the discretization of the differential equation (\ref{dyn.eqn}) was done with a consistent finite difference method (standard second-order finite differences), and the time step was chosen following the Courant-Friedrichs-Lewy stability criterion, $\Delta t\le |k|\ \Delta x$, where $k$ is some constant, we can thus assume that the algorithms used are convergent \cite{bertsekas}. In addition, during the time evolution, we checked the value of the free energy evolution in time to see how it relaxes to a stationary value in a monotonically decreasing way. The values of the inner volume and the surface area were also computed during the evolution and it can be seen that the volume remains constant (up to the numerical precision) during all the process, and similarly with the surface area (the value of the Lagrange multiplier converges rapidly to the stationary solution).

\section{\label{sec:results}Results}

In our model there seem to be several free parameters ($\epsilon$, $a_0(x)$, $V_0$). However, $\epsilon$ is a small parameter (the model is shown to be robust under variations of this parameter), which will be set, in what follows, to be equal to the mesh size.

In addition, the term proportional to $\bm{\nabla}\bar{\sigma}(\bm{x})$ in the dynamic equation (the last term in Eq.~(\ref{dyn.eqn})) is shown to be small, and the Lagrange multiplier, $\bar{\sigma}$, can be considered homogeneous. To show this, we numerically compute these variations for an initial profile of $\bar{\sigma}(x)$, and see how they rapidly relaxe to a nearly constant function (i.e the time scale of the relaxation of the effective surface tension is smaller than the time scale related to the shape change). Moreover, $\sigma(\bm{x})$ appears as an effective surface tension which prevents the surface area from changing. Anyway, its value in membranes is very small compared with other energy scales in the system (e.g. bending rigidity) \cite{evans90}. Therefore, its variations are also small.

Finally, scale invariance causes that the ratio between the constrained total volume and the total surface area is the only relevant parameter in the model (for a fixed topology). Thus, we define a dimensionless volume $v$ as the ratio between the actual volume and the volume of a sphere with the same area,
\eq\label{red.vol}
v=\frac{V}{(4\pi/3) R_0^3},
\eqq
where
\eq
R_0=\left(\frac{A}{4 \pi}\right)^{1/2}.
\eqq
We will thus look for stationary shapes for different fixed topologies: spherical (Euler characteristic equal to 2) and non-spherical (e.g. genus-1 toroidal topologies with Euler characteristic equal to 0). We will focus mainly on spherical topology, in order to discuss the model and the results obtained.

\subsection{\label{sec:sph.top}Spherical topology}

\begin{figure}[htbp!]
\begin{center}
\subfigure[$v=0.69$]{\includegraphics[width=5.0cm]{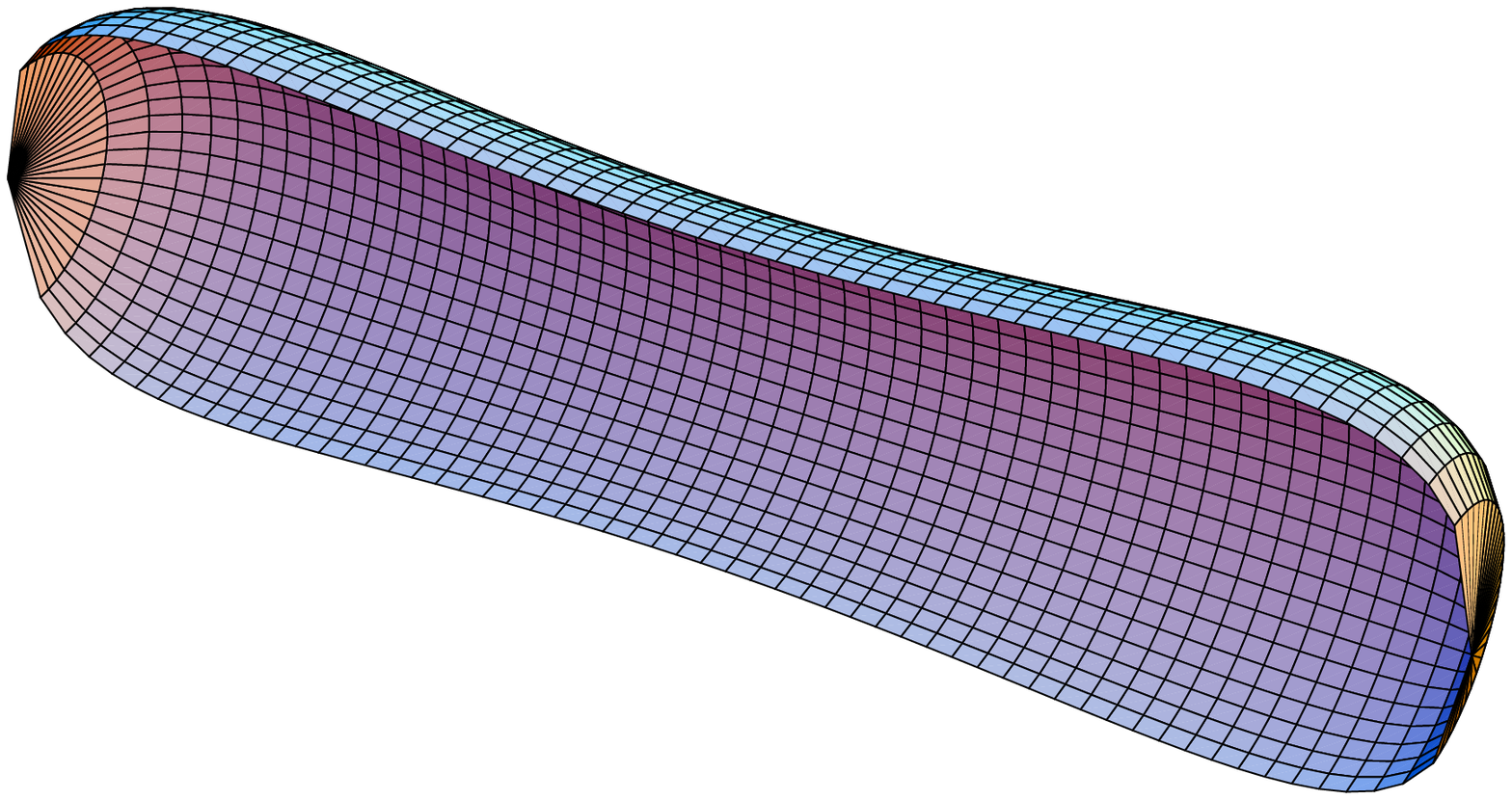}}
\subfigure[$v=0.60$]{\includegraphics[width=5.0cm]{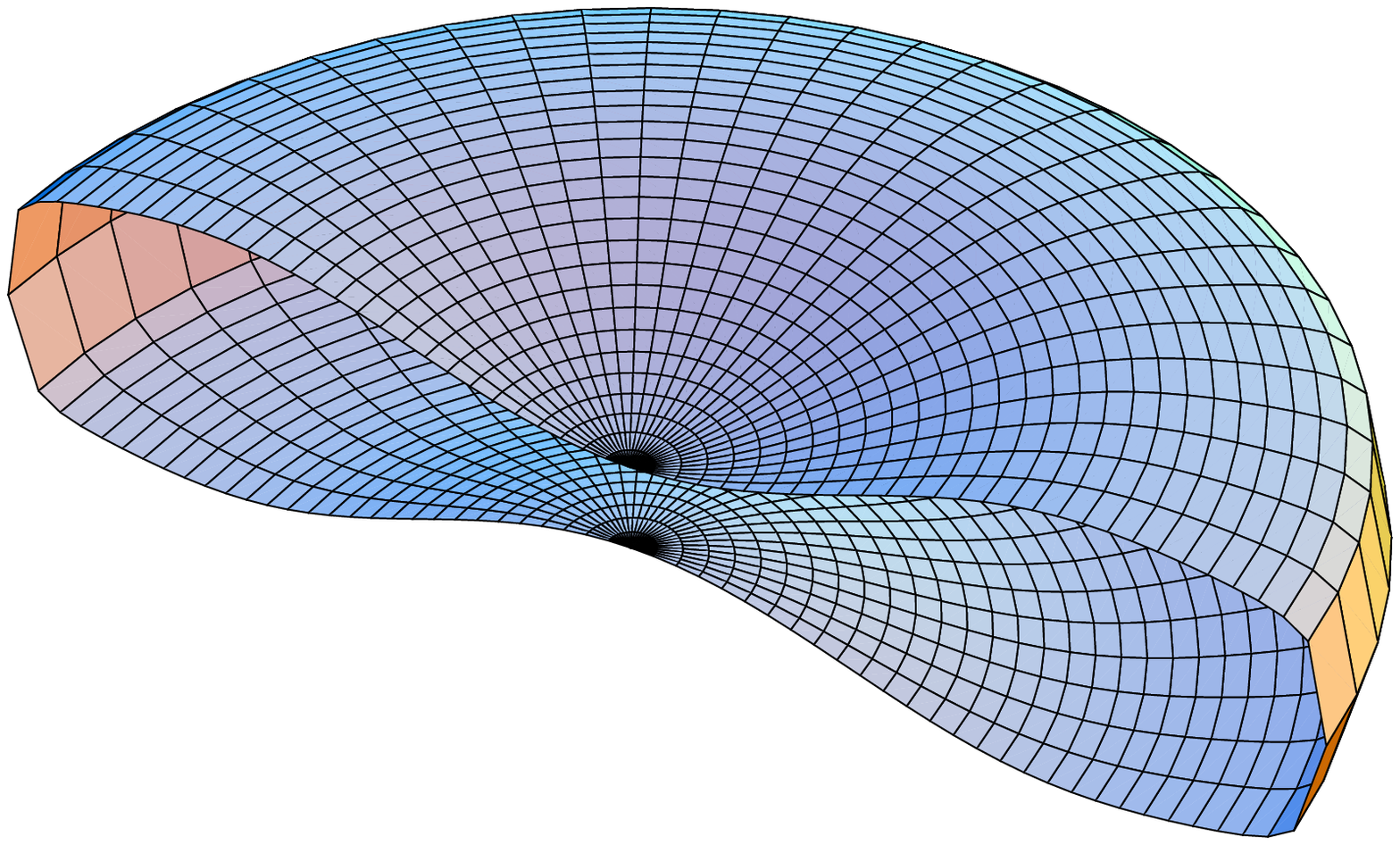}}
\subfigure[$v=0.43$]{\includegraphics[width=3.5cm]{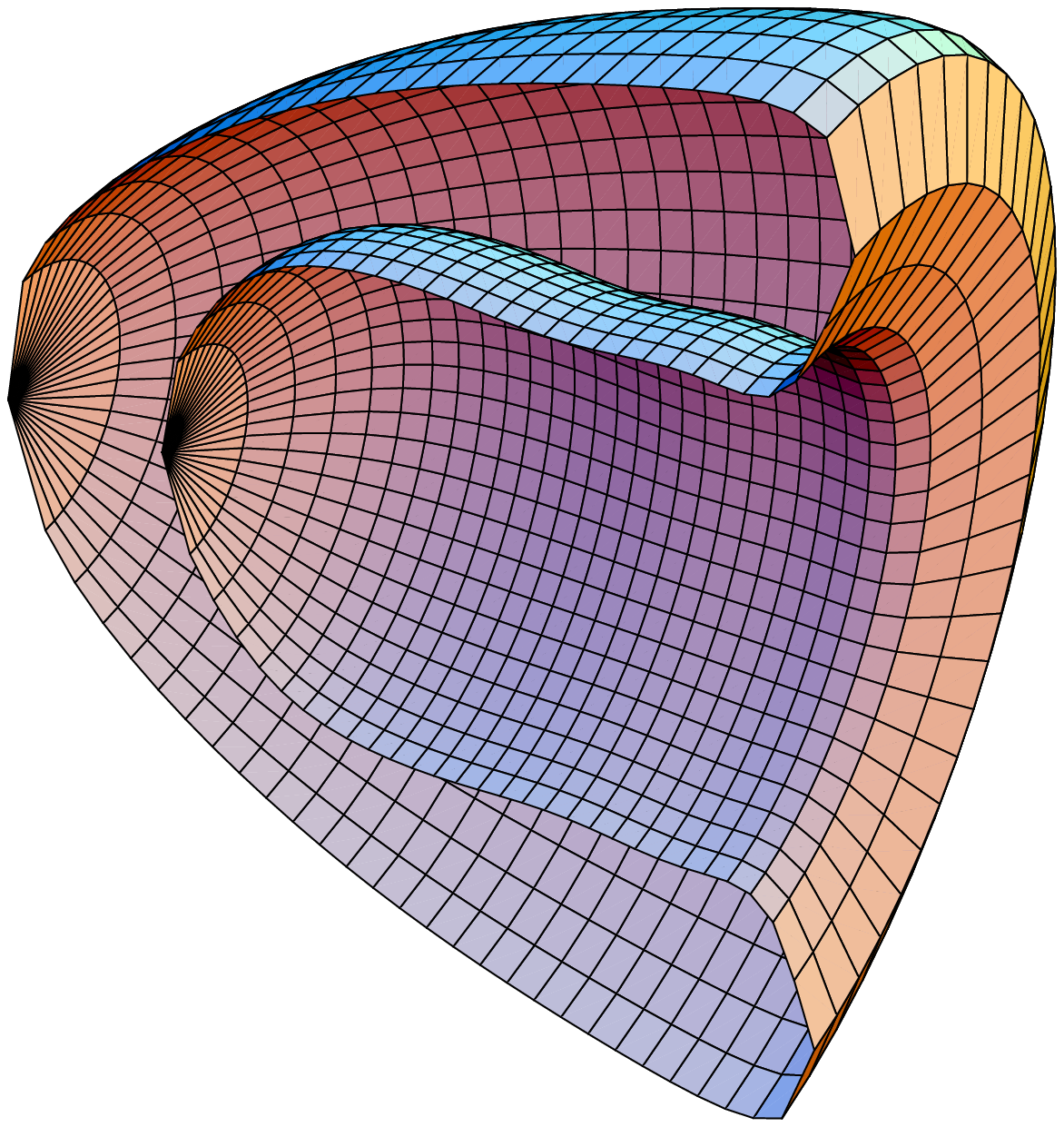}}
\end{center}
\caption{Stationary shapes for the minimal model, as stationary states of the dynamic evolution of certain initial conditions under Eq.~(\ref{dyn.eqn}). (a) Prolate, (b) oblate, and (c) stomatocyte are shown.}
\label{fig:sph.shapes3d}
\end{figure}

For this topology, the three qualitatively different stationary shapes for the minimal model are found, in agreement with \cite{seifert91pra}. These are, in order of decreasing reduced volume, the prolate and oblate ellipsoids and the stomatocyte (see Fig.~\ref{fig:sph.shapes3d}).

Shape evolution is done starting from an arbitrary initial shape. Since the dynamic equation has no external noise (just numerical noise), we start from different initial shapes corresponding to each value of the reduced volume. The initial condition for the phase-field is a sharp distribution of $\phi=+1$ and $\phi=-1$. There is thus a transient period in the first few steps of the evolution, where the diffuse interface is created and a tanh-like profile is obtained, which remains during the subsequent evolution. This transient is needed to calculate the surface area using Eq.~(\ref{globalarea}), since a gradient in $\phi$ is required.

\begin{figure}[htbp!]
\subfigure[$10^4\ \Delta t$]{\includegraphics[width=4cm]{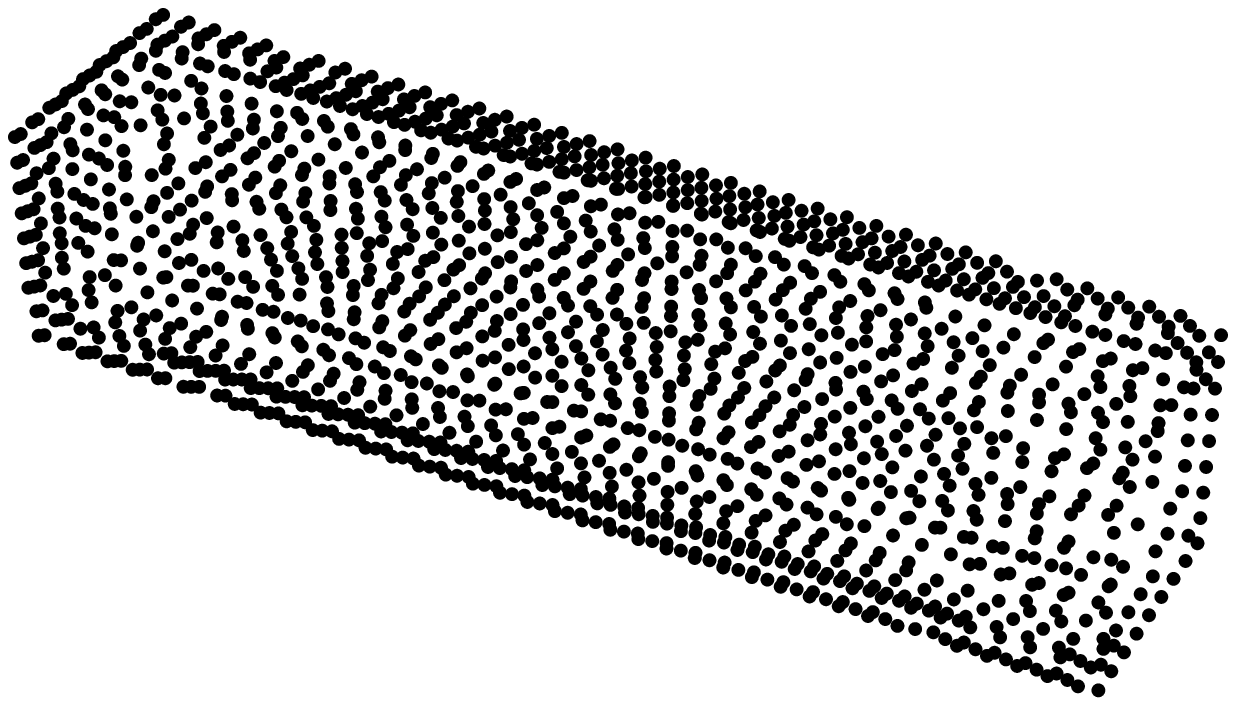}}
\subfigure[$0.5\times 10^6\ \Delta t$]{\includegraphics[width=4cm]{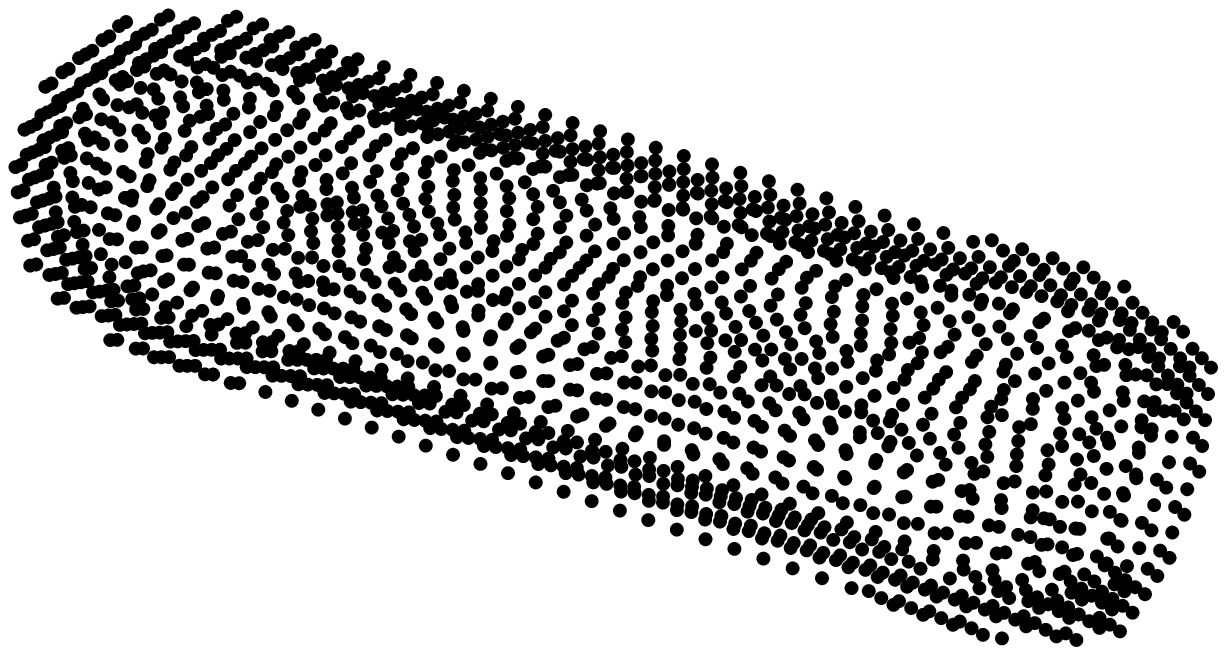}}
\subfigure[$2\times 10^6\ \Delta t$]{\includegraphics[width=4cm]{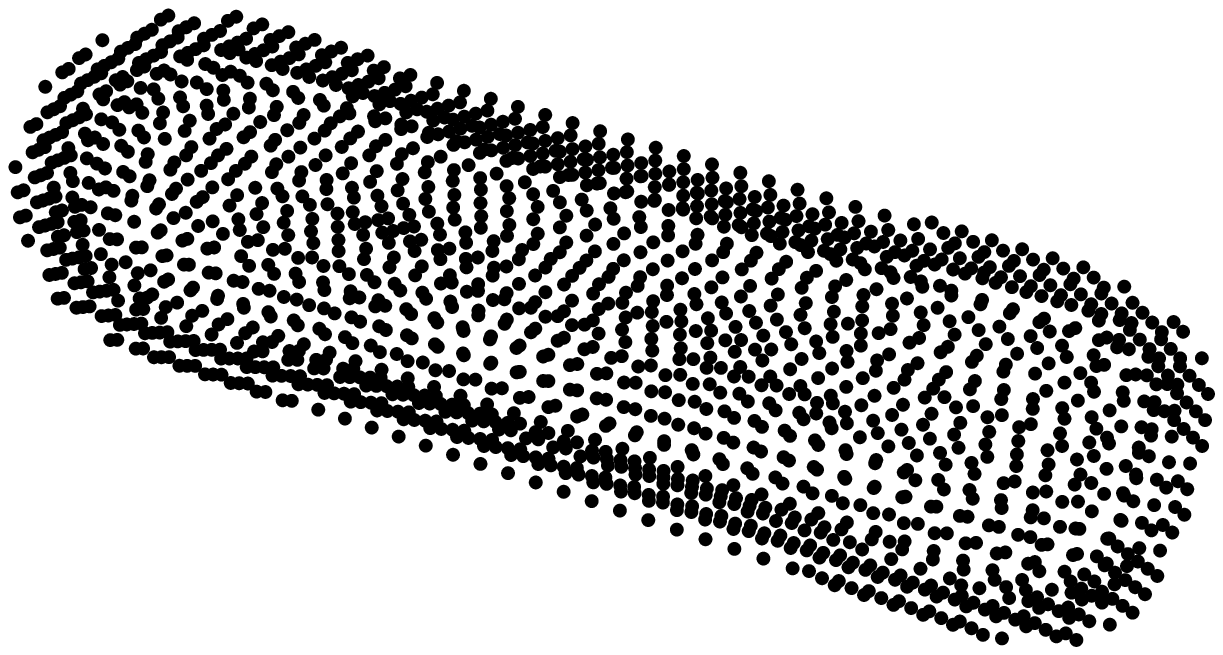}}
\subfigure[$8\times 10^6\ \Delta t$]{\includegraphics[width=4cm]{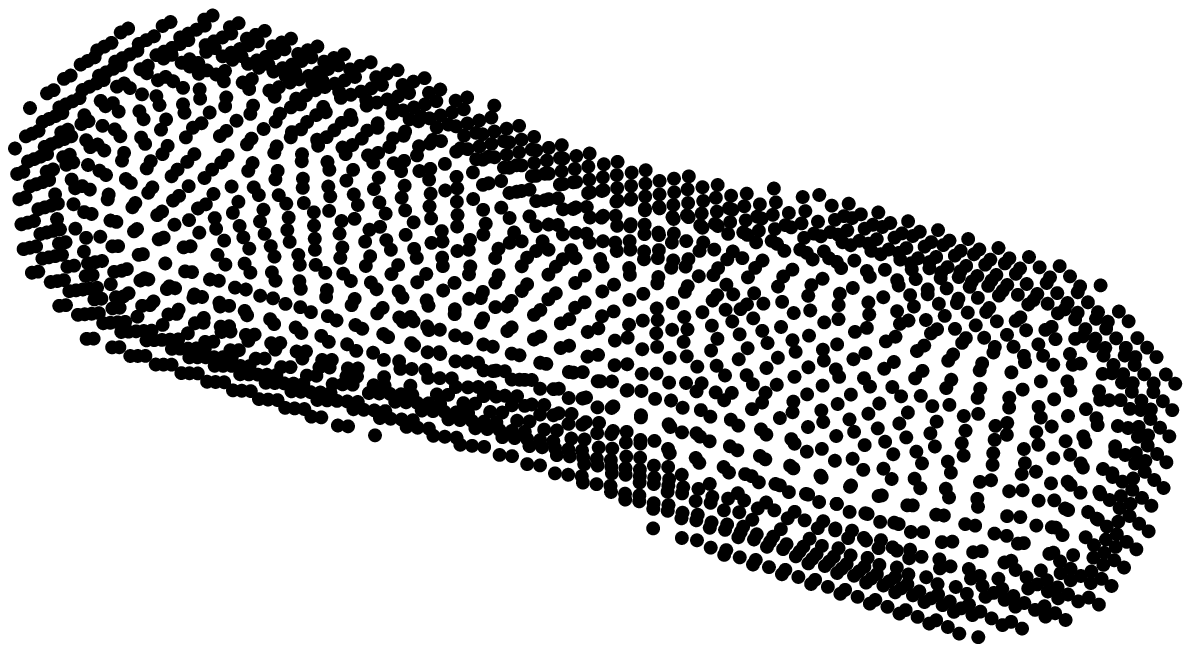}}
\caption{Shape evolution for a vesicle with $v=0.69$, which eventually reaches a prolate shape. the integration was performed on a 3-dimensional $50\times 50\times 50$ lattice with time step $\Delta t=10^{-4}$. No axis of symmetry is supposed here, and the initial shape is a non-axisymmetric $40 \times 10 \times 10$ box.}
\label{fig:prolate3d}
\end{figure}

We have solved Eq.~(\ref{dyn.eqn}) numerically on a three-di\-men\-sion\-al lattice. The possibility of finding non-axisymmetric shapes then arises. In Fig.~\ref{fig:prolate3d}, four snapshots of the shape evolution towards a prolate ellipsoid with $v=0.69$ are shown. This is a stable shape, since the actual transition between prolates and oblates happens at a value of the reduced volume that is lower than this value ($v_D\simeq 0.65$) \cite{seifert91pra}. The dynamic evolution towards this axisymmetric prolate is done with a non-axisymmetric dynamics, and no axis of symmetry has been supposed. The initial shape (Fig.~\ref{fig:prolate3d}(a)) is a non-axisymmetric box, which dynamically evolves towards an axisymmetric shape. 

\begin{figure}[htbp!]
\centering
\setlength{\fboxsep}{0cm}
\subfigure[$10^3\ \Delta t$]{\fbox{\includegraphics[width=4.2cm]{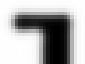}}}
\subfigure[$0.5\times 10^7\ \Delta t$]{\fbox{\includegraphics[width=4.2cm]{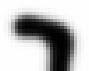}}}
\subfigure[$10^7\ \Delta t$]{\fbox{\includegraphics[width=4.2cm]{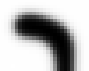}}}
\subfigure[$8\times 10^7\ \Delta t$]{\fbox{\includegraphics[width=4.2cm]{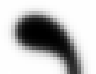}}}
\caption{Shape evolution for a vesicle with $v=0.43$, which eventually reaches a stomatocyte shape. A two-dimensional section is shown, where an axis of symmetry exists, located on the lower side of each snapshot. Integration was performed on an axisymmetric $60\times 30$ lattice and the time step was $\Delta t=10^{-4}$.}
\label{fig:stoma2d}
\end{figure}

We have also studied the behavior of the dynamic equation in the axisymmetric case, where we can discretize Eq.~(\ref{dyn.eqn}) on a two-dimensional lattice. Fig.~\ref{fig:stoma2d} shows the time evolution to eventually reach a stomatocyte-like shape. Plots show the value of the phase-field on a grey-level scale, where black represents $\phi=+1$, or the inner volume of the vesicle; and white corresponds to $\phi=-1$, or the outer volume of the vesicle.

\begin{figure}[htbp!]
\subfigure[$1.1\times 10^8\ \Delta t$; $v\simeq 0.494$]{\includegraphics[width=4.2cm]{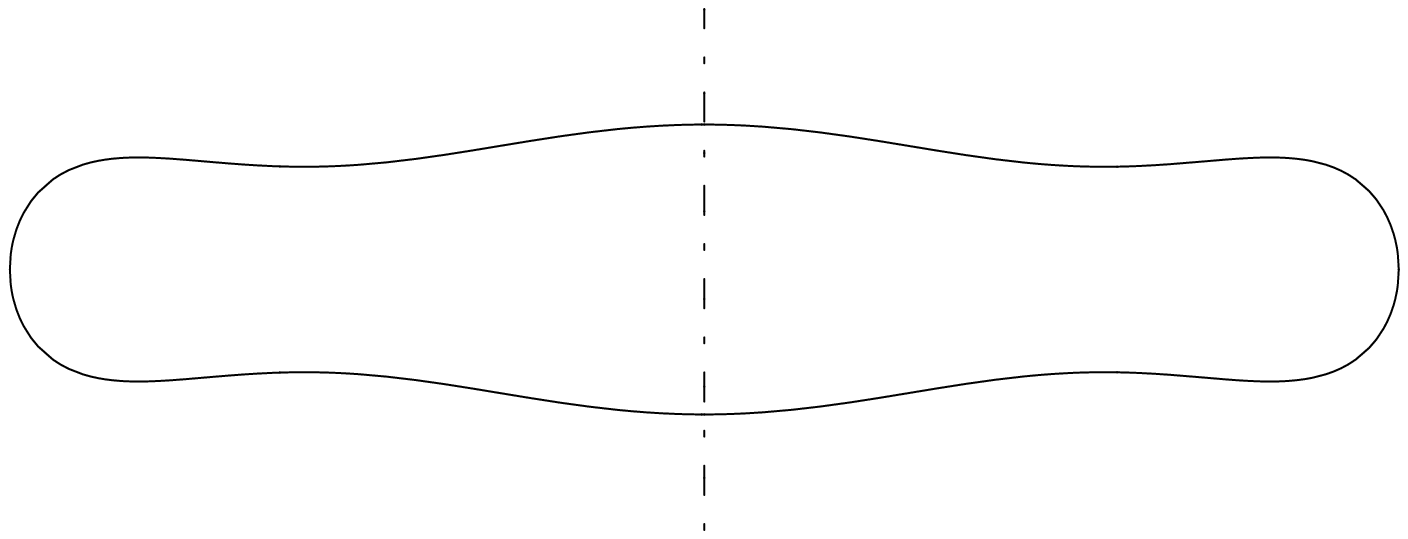}}
\subfigure[$2.2\times 10^8\ \Delta t$; $v\simeq 0.493$]{\includegraphics[width=4.2cm]{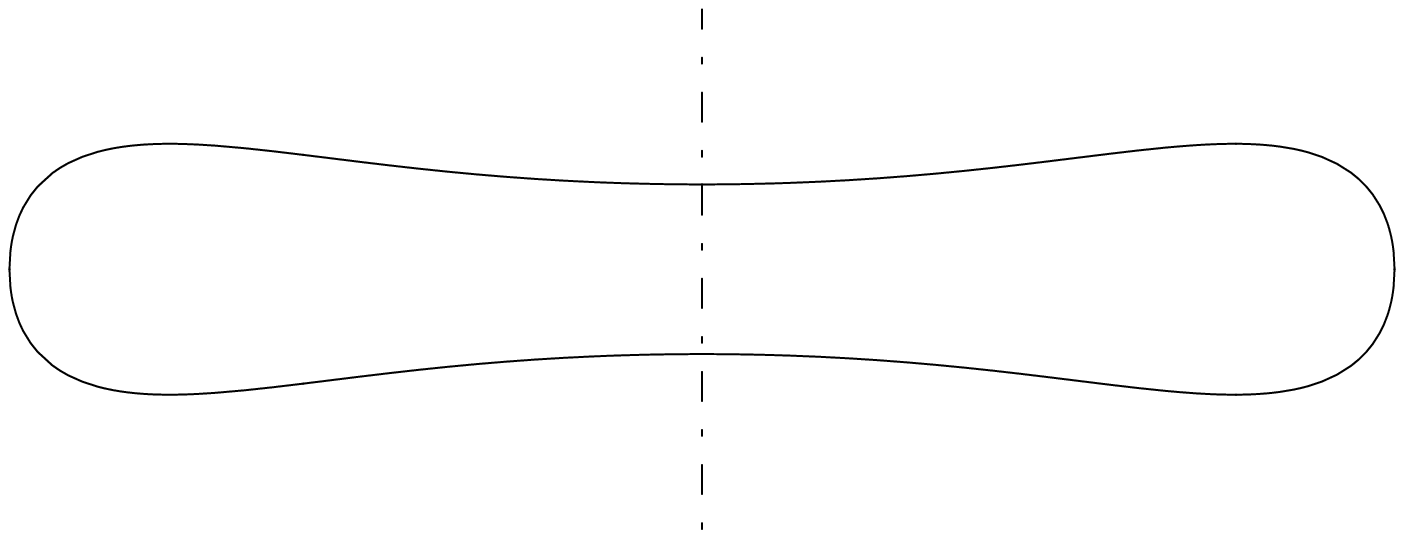}}
\subfigure[$3.3\times 10^8\ \Delta t$; $v\simeq 0.492$]{\includegraphics[width=4.2cm]{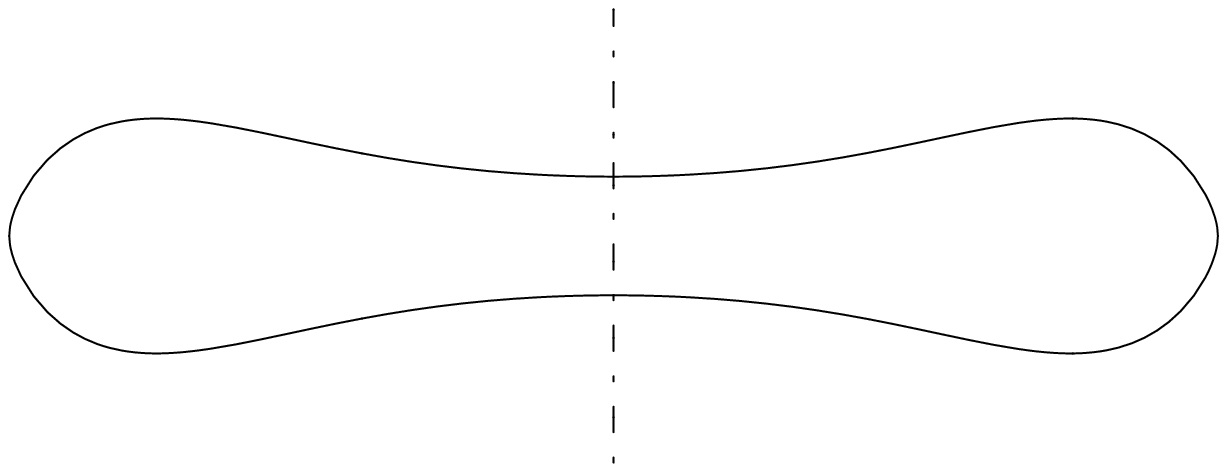}}
\subfigure[$7.8\times 10^8\ \Delta t$; $v\simeq 0.496$]{\includegraphics[width=4.2cm]{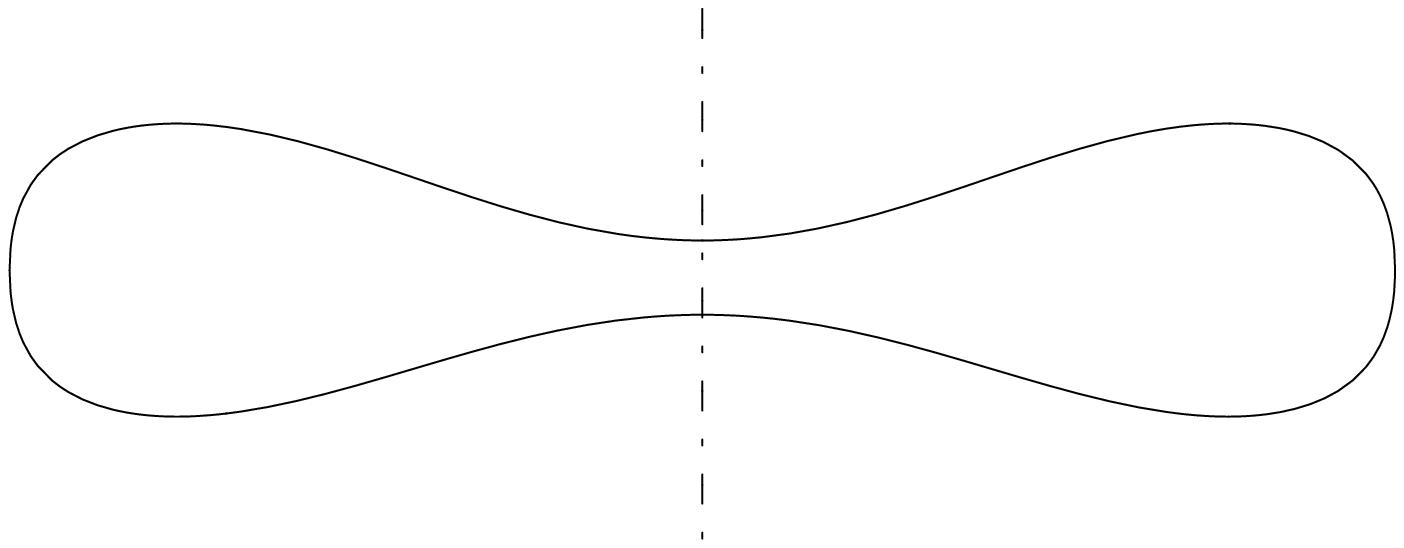}}
\caption{Snapshots of the time evolution towards a discocyte-like shape. Figures shown here are for the axisymmetric case. The dot-dashed line indicates the axis of symmetry. Curves are continuous fits of the level-set of the phase-field on a $80\times 40$ lattice. The time step is set to $\Delta t=10^{-4}$.}
\label{fig:oblate2da}
\end{figure}

We have also found the level-set of these grey-level scale plots, to track the position of the vesicle membrane. In Fig.~\ref{fig:oblate2da} a continuous fit of this contour is plotted and the evolution towards a discocyte-like shape is shown. We can see that the reduced volume corresponding to this figure is $v\simeq 0.5$, which is conserved during evolution (so they are the surface area and the volume separately). This shape is known to be metastable. However, this shape is obtained from the initial ellipsoid because the actual stable shape, corresponding to that value of the reduced volume, is far away in the shape landscape from our initial choice.
Note also that the dynamic equation (\ref{dyn.eqn}) is such as the free energy (\ref{ansatz}) is a monotonically decreasing function, which reaches metastable or stable states where the value of the energy remains constant.

\subsection{\label{sec:tor.top}Non-spherical topologies}

\begin{figure}[htbp!]
\begin{center}
\subfigure[$v=0.44$]{\includegraphics[width=4cm]{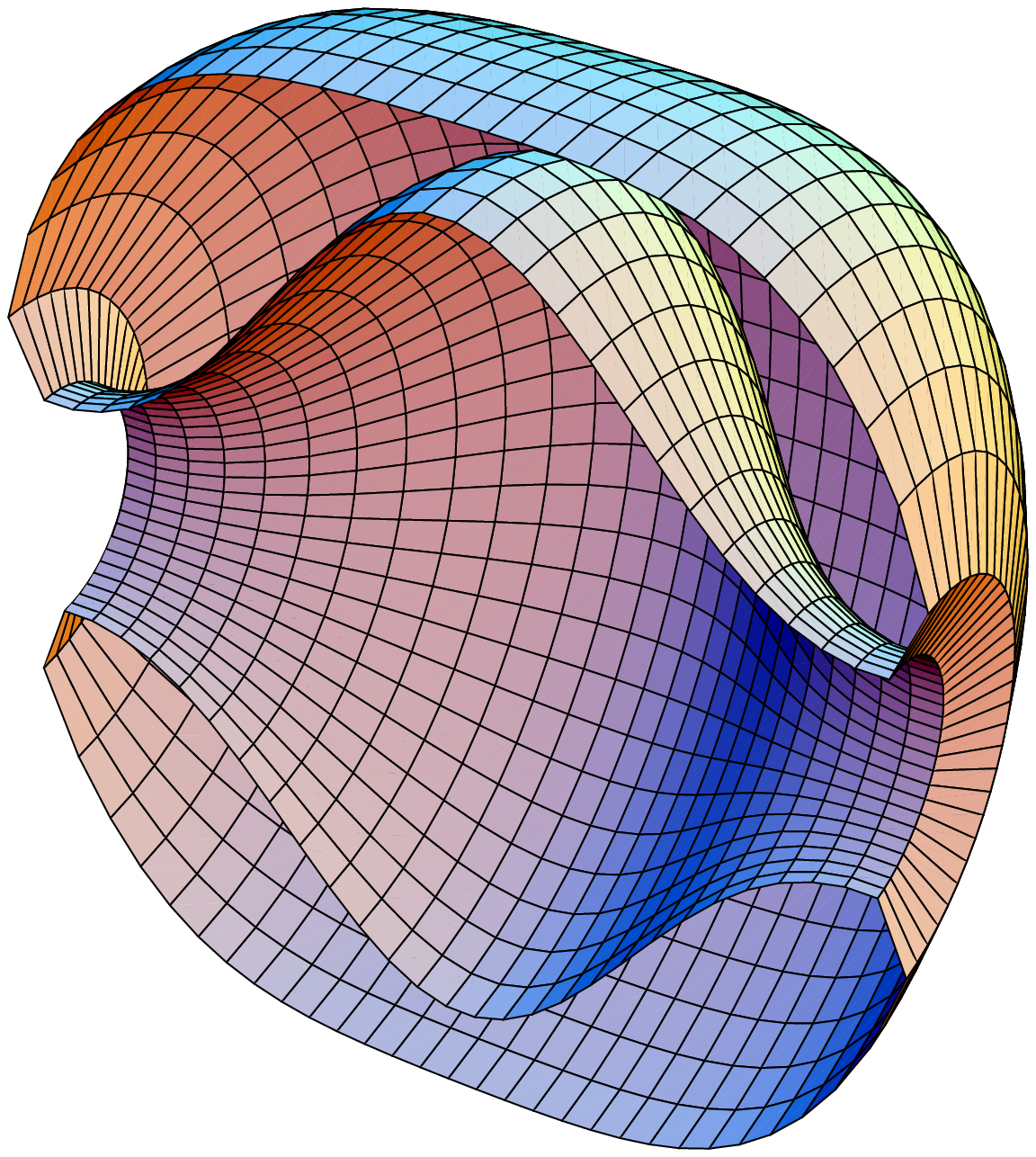}}
\subfigure[$v=0.71$]{\includegraphics[width=4.5cm]{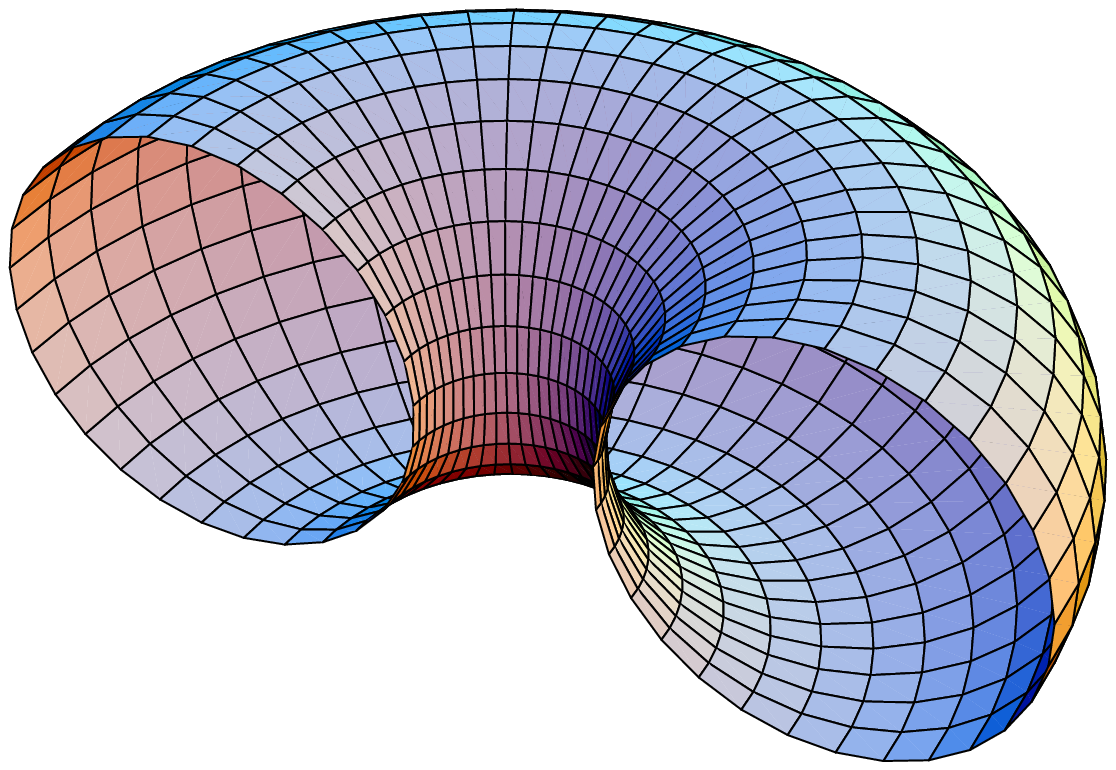}}
\subfigure[$v=0.27$]{\includegraphics[width=4cm]{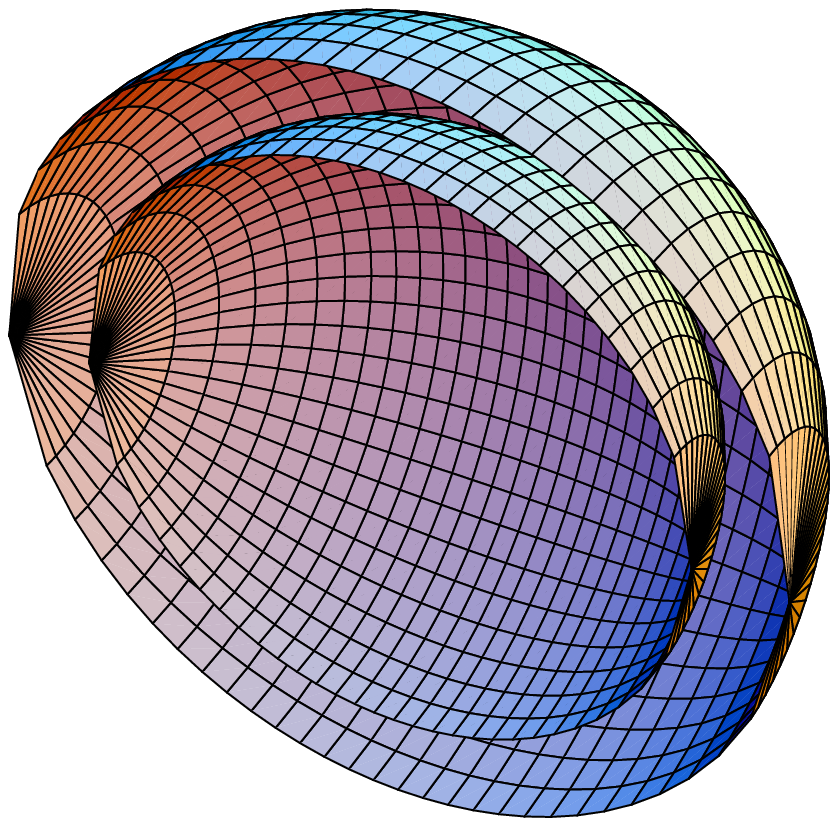}}
\end{center}
\caption{Stationary shapes for non-spherical topologies, as stationary states of the dynamic evolution of certain initial conditions under Eq.~\ref{dyn.eqn}. (a) Sickle-shaped torus of genus-1 toroidal topology, (b) circular toroid (Clifford torus \cite{seifert91prl}) and genus-1 toroidal topology, and (c) spherical shell with an Euler characteristic $\chi=4$ are shown.}
\label{fig:tor.shapes3d}
\end{figure}

\begin{figure}[htbp!]
\centering
\setlength{\fboxsep}{0cm}
\subfigure[$10^3\ \Delta t$]{\fbox{\includegraphics[width=4.2cm]{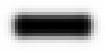}}}
\subfigure[$10^7\ \Delta t$]{\fbox{\includegraphics[width=4.2cm]{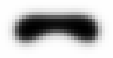}}}
\subfigure[$5\times 10^7\ \Delta t$]{\fbox{\includegraphics[width=4.2cm]{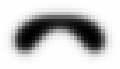}}}
\subfigure[$6.8\times 10^7\ \Delta t$]{\fbox{\includegraphics[width=4.2cm]{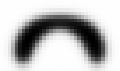}}}
\subfigure[$7\times 10^7\ \Delta t$]{\fbox{\includegraphics[width=4.2cm]{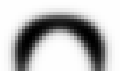}}}
\subfigure[$8\times 10^7\ \Delta t$]{\fbox{\includegraphics[width=4.2cm]{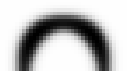}}}
\caption{Shape evolution for a vesicle with $v=0.27$, which eventually reaches a spherical shell shape. A two-dimensional section is shown, where an axis of symmetry exists, located on the lower side of each snapshot. Integration was performed on an axisymmetric $50\times 25$ lattice and the time step was $\Delta t=10^{-4}$. The initial shape (a) has a genus-1 toroidal topology. It evolves dynamically untill it changes its topology towards an spherical shell.}
\label{fig:shell2d}
\end{figure}

We have also studied non-spherical topologies, such as the genus-1 toroidal topology \cite{seifert91prl,julicher93jph} (see Fig.~\ref{fig:tor.shapes3d}(a) and (b)). Circular tori are found for large values of the reduced volume $v$, and sickle-shaped tori for small values of $v$. Discoids can be found for intermediate values of $v$, although they are not stable shapes, and they will eventually fall to stable ones. In addition, spherical shells are also found (see Fig.~\ref{fig:tor.shapes3d}(c)). They have a different topology with an Euler characteristic $\chi=4$. These shapes can be thought of as a limit case of a sickle-shaped torus, when the outer radius vanishes thus changing the global topology of the shape (Fig.~\ref{fig:shell2d}).

The shapes shown in Fig.~\ref{fig:tor.shapes3d} may be found in different ways. First, it is possible to take an initial shape of a given topology, and let it relax to a stationary shape of that same topology. Within this relaxation there is no topological change, and a non-spherical stationary shape may be found. Second, in some cases, dynamic evolution within a given topology leads to shapes close to a topological transition. Then, because of the natural way of dealing with topological transitions of phase-field models, topology may change. We show an example of how we get the shape in Fig.~\ref{fig:tor.shapes3d}(c) in this way in Fig.~\ref{fig:shell2d}. In this Figure there is a topological transition between steps \ref{fig:shell2d}(d) and \ref{fig:shell2d}(e). The poles of the torus get closer to the axis of symmetry and, eventually, fuse to get spherical-shell-like topology. The actual dynamics of this transition is not explained by our model, neither by energetics of the Canham-Helfrich model, since the process of membrane fusion is far from being well-understood. When membranes are fused or broken, the Helfrich approach is not valid, since the microscopic details of the bilayer become then relevant (namely, the radii of curvatures involved are of the order of the membrane width). In addition, when a topological change occurs, there is an energy change due to the Gaussian curvature term in Eq.~(\ref{localbending}).

In any case, once the shape has changed its topology the evolution leads the shape to a stationary state given that new topology (we can think of the initial shape being the shape after the topological change).

%

\section{\label{sec:discussion}Shape Diagram}

\begin{figure}[htbp!]
\centering
\vskip5mm
\includegraphics[angle=-90,width=9.8cm,bb=100 50 612 792]{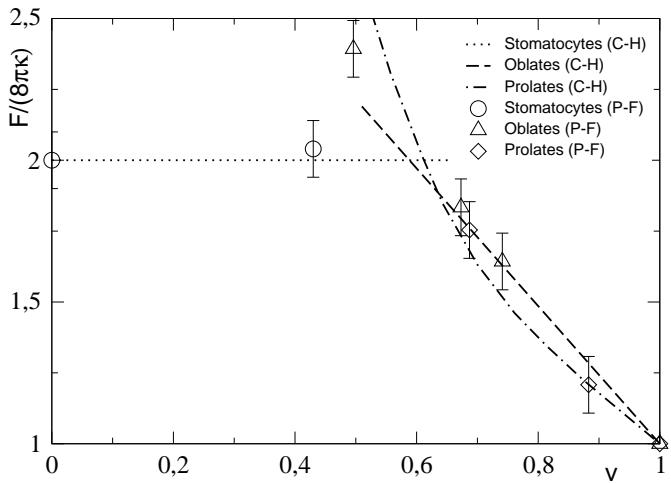}
\caption{Shape diagram for the minimal model with spherical topology. Lines correspond to the minimization of the Canham-Helfrich energy (C-H) \cite{seifert91pra}, and symbols to the results of the phase-field model (P-F).}
\label{fig:phase.diag}
\end{figure}

The results shown in the last section are collected in a shape diagram where the bending energy is plotted against the reduced volume (see Fig.~\ref{fig:phase.diag}). The curvature energy of the shapes obtained as stationary states of the dynamic evolution under Eq.~(\ref{dyn.eqn}), is calculated in the following way. Firstly, the interface is located as the level-set, $\phi=0$, of the phase-field. An interpolation is then performed over the discrete data in order to obtain a continuous function describing the membrane. Using surface differential geometry \cite{docarmo76,safran94}, the curvature tensor on the surface defined by the interpolating function rotated about the axis of symmetry is worked out. The trace of this tensor, related with the mean curvature, is then calculated. Integrating over the surface eventually gives the bending energy.


Once these pairs $\{v,F\}$ are obtained, a comparison with known results is presented. Thus, in Fig.~\ref{fig:phase.diag}, we plot the lines corresponding to the minimization of the Canham-Helfrich free energy, Eq.~(\ref{globalbending}), with fixed volume and surface area \cite{seifert91pra}. There are three branches of different shapes (stomatocytes, oblates and prolates), which intersect at certain values of the reduced volume, $v$, where a change in the stability of the shape occurs. We see that the results obtained for stationary shapes with the phase-field Eq.~(\ref{ansatz}) are in good agreement with this.

\section{\label{sec:conclusions}Discussion and Conclusions}

The model could be extended to include hydrodynamic effects. Since in this paper we give a derivation of the Canham-Helfrich free energy within a phase-field approach, a hydrodynamic equation for the velocity field (e.g. the Navier-Stokes equation) could be introduced, with a force acting on the membrane due to the bending elasticity. In addition, an advective term should be included in the dynamic equation for the phase-field \cite{misbah03}. However, our future aim is to study the shape instabilities seen in \cite{tsafrir01,tsafrir03}. There, the relevant characteristic time scale that needs to be studied is that associated with the relaxation of the curvature, which turns out to be related to the diffusion coefficient of the polymer in the membrane, and not directly to the membrane viscosity. Therefore, a dynamic model which couples the polymer concentration in the membrane with the local spontaneous curvature would be needed.

The existence of a differential flow between monolayers (i.e. sliding between both monolayers) due to the viscous drag between the non-polar tails of the hydrocarbon chains of the lipids is a microscopic effect which may induce a velocity discontinuity at the membrane. Thus, it would be an important factor to keep in mind when introducing the hydrodynamic field \cite{seifert93,yeung95}. When dealing with shape instabilities due to the anchorage of amphiphilic molecules \cite{tsafrir01,tsafrir03}, curvature can be induced by two mechanisms, both of which are theoretically explained by a model: the area difference elasticity (ADE) model, and the spontaneous curvature (SC) model \cite{seifert97}. ADE relaxes via the sliding between monolayers, and SC via diffusion of the polymer in the membrane. From experiments, it is possible to distinguish between the ADE and the SC mechanisms by measuring the relaxation times (or the diffusion constants). In the experiments done by Tsafrir \emph{et al.} \cite{tsafrir01}, evidence that both ADE and SC mechanisms influence pearling instability was provided. However, it was assumed there that pearling was a result of the SC mechanism, i.e. that the dynamic instability was not affected by the sliding between the monolayers of the membrane.

As a conclusion, a phase-field model for bending energy of fluid vesicles, and not surface tension as in usual phase-field models, has been derived. It has been shown to be equivalent in the sharp-interface limit to the Canham-Helfrich free energy for closed membranes. Various geometric constraints are also implemented in the phase-field approach. Dynamic equations for the phase-field and for the Lagrange multiplier related to the surface area conservation have been worked out explicitly. These are highly non-linear partial differential equations and thus numerical treatment was found to be necessary. Stationary shapes of vesicles with different topologies are found, in agreement with those obtained by minimization of the Canham-Helfrich free energy \cite{seifert97}. In addition, they are found dynamically, from arbitrary initial shapes. Moreover, a shape diagram for spherical topology vesicles is presented. These facts show that our phase-field model is a good description of the bending elasticity of membranes, and that it could be used within a generalized dynamic framework where hydrodynamics or polymer diffusion on the membrane could be introduced.

\begin{acknowledgements}

We are grateful to Joel Stavans for drawing our attention to the problem of membranes. We acknowledge financial support of the Direcci\'{o}n General de Investigaci\'{o}n under project No. BFM2003-07749-C05-04. F.C. also thanks Ministerio de Educaci\'{o}n y Ciencia (Spain) for financial support.

\end{acknowledgements}

\end{document}